# Do You Always Need a Textbook to Teach Astro 101?

Alexander L. Rudolph

Department of Physics and Astronomy, California State Polytechnic University, Pomona, CA 91768

ABSTRACT

The increasing use of interactive learning [IL] strategies in Astro 101 classrooms has led some instructors to consider the usefulness of a textbook in such classes. These strategies provide students a learning modality very different from the traditional lecture supplemented by reading a textbook and homework, and raises the question of whether the learning that takes place during such interactive activities is enough by itself to teach students what we wish them to know about astronomy.

To address this question, assessment data is presented from an interactive class, which was first taught with a required textbook, and then with the textbook being optional. Comparison of test scores before and after this change shows no statistical difference in student achievement whether a textbook is required or not. In addition, comparison of test scores of students who purchased the textbook to those who did not, after the textbook became optional, also show no statistical difference between the two groups. The Light and Spectroscopy Concept Inventory (LSCI; Bardar et al. 2007), a research-validated assessment tool, was given pre- and post-instruction to three classes that had a required textbook, and one for which the textbook was optional, and the results demonstrate that the student learning gains on this central topic were statistically indistinguishable between the two groups. Finally, the Star Properties Concept

Inventory (SPCI), another research-validated assessment tool, was administered to a class for which the textbook was optional, and the class performance was higher than that of a group of classes in a national study (Bailey et al. 2011).

Taken together, these results suggest that, if research-tested alternative learning modalities are provided, students' ability to learn the content of an Astronomy 101 course can be independent of a textbook requirement. Details on the course and the methodology used to reach this conclusion are presented.

## I. INTRODUCTION

Modern Astro 101 textbooks are things of beauty. They are full of up-to-date information on all types of astronomical phenomena, presented in a visually stunning style. They often incorporate best pedagogical practices in the delivery of material, and come packaged with extensive supplementary material for both the instructor and student. On-line homework and study sites provide additional material, including real-time feedback and hints in homework problems and in tutorial-style activities.[1] During educational sessions at astronomy meetings, one of the questions frequently discussed is which textbook to use. Some instructors have a favorite, while others will claim that they are all the same, and that therefore it does not matter. Increasingly, the discussions turn to the somewhat radical question of whether any textbook is needed at all.

---

[1] See Bruning (2007) for a recent survey of Astro 101 textbooks.

Early research into science education showed that, not surprisingly, textbooks play a central role in the teaching of science. In fact, many of these studies showed that the textbook is often the primary source of knowledge for the student, and even dictates the curriculum (Gottfried & Kyle 1992). Instructors, particularly early in their careers, often use the textbook as a guide to creating their syllabus.

The seminal work *How People Learn: Brain, Mind, Experience, and School* (National Research Council 1999) discusses the role textbooks have in helping students learn. It notes that research shows that students (novices), unlike experts in a field, lack what is termed "conditional knowledge", namely the ability to place factual knowledge in the context in which it is useful. It is this conditional knowledge that allows a student to "know when, where, and why to use that knowledge" (p. 37). It also notes that textbooks often emphasize factual, or declarative, knowledge, and do not usually give students the tools needed to "conditionalize" that knowledge. "Textbooks are much more explicit in enunciating the laws of mathematics or of nature than in saying anything about when these laws may be useful in solving problems" (p. 31).

*How People Learn* goes on to detail research showing that students enter the classroom with firmly entrenched preconceptions which, if not confronted, will limit students' ability to accept and synthesize the new knowledge presented to them in a class. It suggests that to overcome these preconceptions, it is necessary to confront them by creating "conceptual conflict", the realization that an intuitive preconception, often based on everyday experiences, is in conflict with the scientific principles being presented, a process known as conceptual change. (For a description of a leading theory of conceptual change, see Posner et al. 1982.)

Most traditional textbooks use a relatively straightforward, clear, well-written exposition of accepted scientific concepts, described by Guzzetti et al. (1997) as a "non-refutational expository text (informational text that presents only the scientifically acceptable concepts)". A meta-analysis of 25 studies showed that these traditional textbooks were ineffectual at creating either short-term or long-term conceptual change (Guzzetti et al. 1993). These findings have led to the development of a new kind of text-based instructional material, known as refutational text, "text that contrasts common alternative conceptions with scientific conceptions" (Guzzetti et al. 1997). Guzzetti et al. (1997) show that refutational text does promote cognitive conflict in students' minds, but go on to state that such cognitive conflict is necessary but not sufficient for conceptual change. Sadler & Tai (2001) found, in a study of almost 2000 students at 18 colleges and universities enrolled in introductory college physics classes, that students who had rigorous high school physics classes that did *not* use a textbook performed statistically significantly better than those in classes that did use a textbook, particularly if their high school teacher concentrated on fewer concepts, covering less material, but in greater depth.

Knight (2004) in his guidebook to physics teaching entitled *Five Easy Lessons: Strategies for Successful Physics Teaching*, reviews a great deal of the literature in Physics Education Research [PER] on conceptual change. He draws the distinction between the "transmissionist model", in which the instructor (or textbook) transmits knowledge to the student, and "scientific constructivism", in which students construct models based on a variety of inputs that might include: textbook reading, lecture, interactive learning activities, homework, laboratories, etc. (McDermott 1991; Redish 1994). To quote Knight, "Active engagement is the essence of the constructivist approach, because students must *build* their mental models rather than receive them from the instructor" (p. 42). He then goes on to list the types of interactive learning strategies that have been shown by research to effectively promote the kind of conceptual change required for students to achieve a deeper conceptual understanding than is typically possible in the traditional classroom, noting, "The common theme is that students are engaged in *doing* or *talking about* physics, rather than listening to physics" (p. 42). The key point is that textbooks fall into the category of teaching modalities that provide declarative knowledge, which has a limited impact on creating conceptual change, and that it is through active engagement with the material, particularly through discourse, that deeper learning occurs.

Motivated by this understanding of student learning, there has been an increasing use of interactive learning strategies such as Think-Pair-Share questions, Lecture-Tutorials, and Ranking Tasks (Crouch & Mazur 2001; Prather et al. 2004; Hudgins et al. 2006; Prather et al. 2008) in Astro 101 classes. Among Astro 101 students, there is evidence that such strategies lead to greater conceptual learning than traditional lecture-based classes (Prather et al. 2009; Prather, Rudolph, & Brissenden 2009; Rudolph et al. 2010; Bailey et al. 2011).

All this leads to the question: how well do students learn in a conceptually rich and challenging Astro 101 class that does not require a textbook? To test this question, data were collected from an Astro 101 class taught by a single instructor, covering the same astronomy topics, sometimes with a textbook required, sometimes with the textbook optional. The course was designed to provide basic factual knowledge through short lectures, coupled with active-learning strategies; a textbook was an integral part of the course design process, as shown in Section II, and was only made optional after a few years of the course being taught by this instructor. Student learning in this course was then measured and compared between classes that did and did not require the textbook.

One key to designing a successful study of this kind is determining how student learning is assessed. Exams are, of course, the main way we assess student learning, and exam scores were used as one such measure. As additional, external, measures of student learning, the Light and Spectroscopy Concept Inventory (LSCI; Bardar et al. 2007), and the Star Properties Concept Inventory, SPCI; Bailey et al. 2011), two research-validated assessment tools designed to test students' understanding of a central set of concepts in the course, were administered pre- and post-instruction, to measure student learning gains of these topics in the class.

In Section II, the settings and participants are described, including details on the development of the course content and the role of interactive learning strategies in the course. In Section III, the study design is described. In Section IV, data are presented that allowed the evaluation of the effect on student learning of making the textbook in the course optional, and in Section V, the implications of the study findings are discussed.

## II. SETTING AND PARTICIPANTS

The data for this study comes from an Astro 101 class taught at California State Polytechnic University (Cal Poly Pomona). Cal Poly Pomona is part of the California State University system, the largest public university in the nation, with over 350,000 undergraduate students enrolled at 23 campuses, of which about 19,000 are enrolled as at Cal Poly Pomona.

Demographic information collected as part of this study showed that the makeup of this Astro 101 class over the five years of the study (gender, ethnicity, socioeconomic status) is statistically indistinguishable from that found in a national study of student demographics of almost 1000 students enrolled in Astro 101 courses nationwide (Rudolph et al. 2010). This same study showed that the students in the national study were statistically indistinguishable from U.S. college students in general.

The class consisted of a typical survey of stars, galaxies, and cosmology, with an emphasis on the application of physical laws to understanding the universe. The major learning goals of the course were 1) to help students understand the nature of science through the topic of astrophysics, 2) to develop students' critical reasoning skills enabling them to solve high-level conceptual problems in astronomy, and 3) to develop students' intrinsic interest in astronomy. At the beginning of the study, the course required a textbook, *The Cosmic Perspective: Stars, Galaxies and Cosmology,* by Bennett et al. (2008; hereinafter referred to as Bennett). The book, like most Astro 101 textbooks, covers the topics typically taught in Astro 101 classes: basic physical principles such as Kepler's Laws, Newton's Laws of Motion and Gravity, and the properties of Light and Spectra, followed by the properties of Stars, Galaxies, and Cosmology (as the book's name implies). In addition to being comprehensive and well written, the text matches well with the stated learning goals of the course by focusing on the application of physical laws to astronomy. The students also purchased the workbook, *Lecture-Tutorials for Introductory Astronomy, Second edition* (Prather et al. 2008).[2]

[2] A third edition is now available.

The syllabus for the course was constructed using the textbook as a guide. As is typical, not all material in the textbook was covered: given the comprehensive nature of most Astro 101 textbooks, it would be impossible to cover the entire book in one term, and one makes choices. The class met twice a week for two hours, for a total of 4 hours of class-time per week. Each class period was divided into two 50-minute segments with a 10-minute break in-between. The in-class time was designed using materials and interactive learning [IL] strategies presented at the Center for Astronomy Education [CAE] teaching excellence workshops: Think-Pair-Share [TPS] questions[3], Lecture Tutorials, and Ranking Tasks. The reader is directed to articles on the development and implementation of Lecture-Tutorials (Prather et al. 2004) and Ranking Tasks (Hudgins et al. 2006) in Astro 101 classes for details on these materials.

When the textbook was required, reading on each topic, along with an on-line reading quiz were assigned before class; once the textbook became optional, the reading was listed in the syllabus, but the reading and associated reading quiz were no longer required. A typical class period was divided between short "mini-lectures" on a given topic, followed by a Lecture Tutorial [LT] on the same topic, coupled with pre-LT and post-LT TPS questions to help the students and instructor assess student learning on that topic. Homework assignments consisted of completing LTs that were not finished in class, and Ranking Tasks selected to reinforce learning on the topics of the class. Appendix A contains the schedule for a recent class showing the topics taught, the reading either assigned or suggested (depending on whether the textbook was required or optional), and the Lecture-Tutorials and Ranking Tasks used in each class. The Ranking Tasks assigned as homework were not listed on the syllabus, but were assigned in class.

[3] For details on the best practice implementation of TPS questions in the Astro 101 classroom, see: http://astronomy101.jpl.nasa.gov/teachingstrategies/teachingdetails/?StrategyID=23.

The course was taught by the same instructor 12 times over a period of 5 years. Of these, three were taught as honors classes to students in the Cal Poly Pomona Honors College, and these classes are not included in the study. Between Spring 2007 and Spring 2011, a format of two midterm exams and one final exam was used, and these exams were the same for the 9 times the course was taught in that period, providing a natural controlled study of student learning. The first five of these nine times the course was taught, from Spring 2007 to Spring 2009, the textbook was required. The next four times, from Fall 2009 to Spring 2011, the textbook was optional. The Lecture Tutorial workbook was required throughout.

Although there is no agreement on which of the many topics covered in most Astro 101 textbooks are most important,[4] the Lecture-Tutorials in *Lecture-Tutorials for Introductory Astronomy* have been designed to cover many of the topics commonly taught, including many included in this course (see Appendix A). These Lecture-Tutorials are developed to address the *same topics,* and the *same learning goals* as most Astro 101 textbooks (including Bennett). To illustrate this, Table 1 shows a sample of the learning goals from Bennett's chapter 5, *Light and Matter*, with the relevant LTs used in this course.

For example, Section 5.5 of Bennett and the Lecture-Tutorial on Doppler Shift both address the cause and nature of the Doppler shift, including the difference between redshift and blueshift, the fact that only radial motion causes Doppler shift, and the relationship between the radial speed of an object and the size of the shift. The main difference between the two is that the textbook provides a detailed exposition on these topics, while the Lecture-Tutorial uses a Socratic-dialogue approach of asking questions, and requires explanations to engage the student actively in grappling with the material.

---

[4] See Partridge & Greenstein (2003) for a discussion of this topic.

Table 1. Learning goals of the course textbook and the relevant Lecture-Tutorial topics

| Learning goals for Chapter 5 of Bennett: Light and Matter | Lecture-Tutorial Topics |
| --- | --- |
| 5.1 Light in Everyday Life<br>• How do we experience light?<br>• How do light and matter interact?<br>5.2 Properties of Light<br>• What is light?<br>• What is the electromagnetic spectrum?<br>5.3 Properties of Matter<br>• What is the structure of matter?<br>• What are the phases of matter?<br>• How is energy stored in atoms? | Electromagnetic (EM) Spectrum of Light |
| 5.4 Learning from Light<br>• What are the three basic types of spectra?<br>• How does light tell us what things are made of?<br>• How does light tell us the temperatures of planets and stars?<br>• How do we interpret an actual spectrum? | Types of Spectra<br>Light and Atoms<br>Blackbody Radiation |
| 5.5 The Doppler Shift<br>• How does light tell us the speed of a distant object?<br>• How does light tell us the rotation rate of a distant object? | Doppler Shift |

## III. STUDY DESIGN

As noted above, the course was taught nine times over a period of four years: five times with a required textbook, for with the textbook optional. The common exams allow a direct comparison of student learning between these two periods. In addition, pre-instruction and post-instruction data were collected for two research-validated concept inventories, the Light and Spectroscopy Concept Inventory (LSCI; Bardar et al. 2007), and the Star Properties Concept Inventory, SPCI; Bailey et al. 2011), to further assess student learning independently of the course materials. The former was given three times during the period of the required textbook, and once when the textbook was optional. The SPCI was only given when the textbook was optional, but the results from this class were compared to those in a national study of the SPCI (Bailey et al. 2011).

The exams were entirely multiple-choice, and were taken and graded using Scantron™ forms. The majority of the questions on the exams were developed independently of the instructor of this course, using research on student learning and common misconceptions to develop questions at a variety of levels, and which tend to emphasize reasoning over factual recall. In many cases, multiple reasoning steps are required to answer a question directly. To illustrate this point, Figure 1 shows a few sample exam questions, each of which requires multiple reasoning steps to answer.

Most of the exam questions have one or more attractive distractors to discourage guessing. Exam scores are used as a proxy for student learning, at least in the aggregate. It is important to note that, just as the course syllabus and in-class pedagogy were designed to help the students learn the same material as covered in the typical Astro 101 textbook, so too the exam questions were testing the same content and concepts. To protect the integrity of the exams, the exam booklets were collected at the end of the exam and were not returned to the students. Students who wish to review their performance on an exam come to office hours to go over the exams with the instructor.

Use the four spectra shown at right for objects A-D, to answer the next two questions. Note that one of the spectra is from an object at rest (not moving) and that the remaining spectra come from objects that are all moving toward the observer. *The left end of each spectrum corresponds to shorter wavelengths (blue light) and the right end of each spectrum corresponds to longer wavelengths (red light).*

1. Which two objects appear to be moving with approximately the same speed?
   a. Objects A and B
   b. Objects B and D
   c. Objects C and D
   d. Objects A and D
   e. They are all moving at the same speed, the speed of light.

2. Which of the four objects A-D is moving with the fastest speed?
   a. Object A
   b. Object B
   c. Object C
   d. Object D
   e. More than one object is moving with the fastest speed.

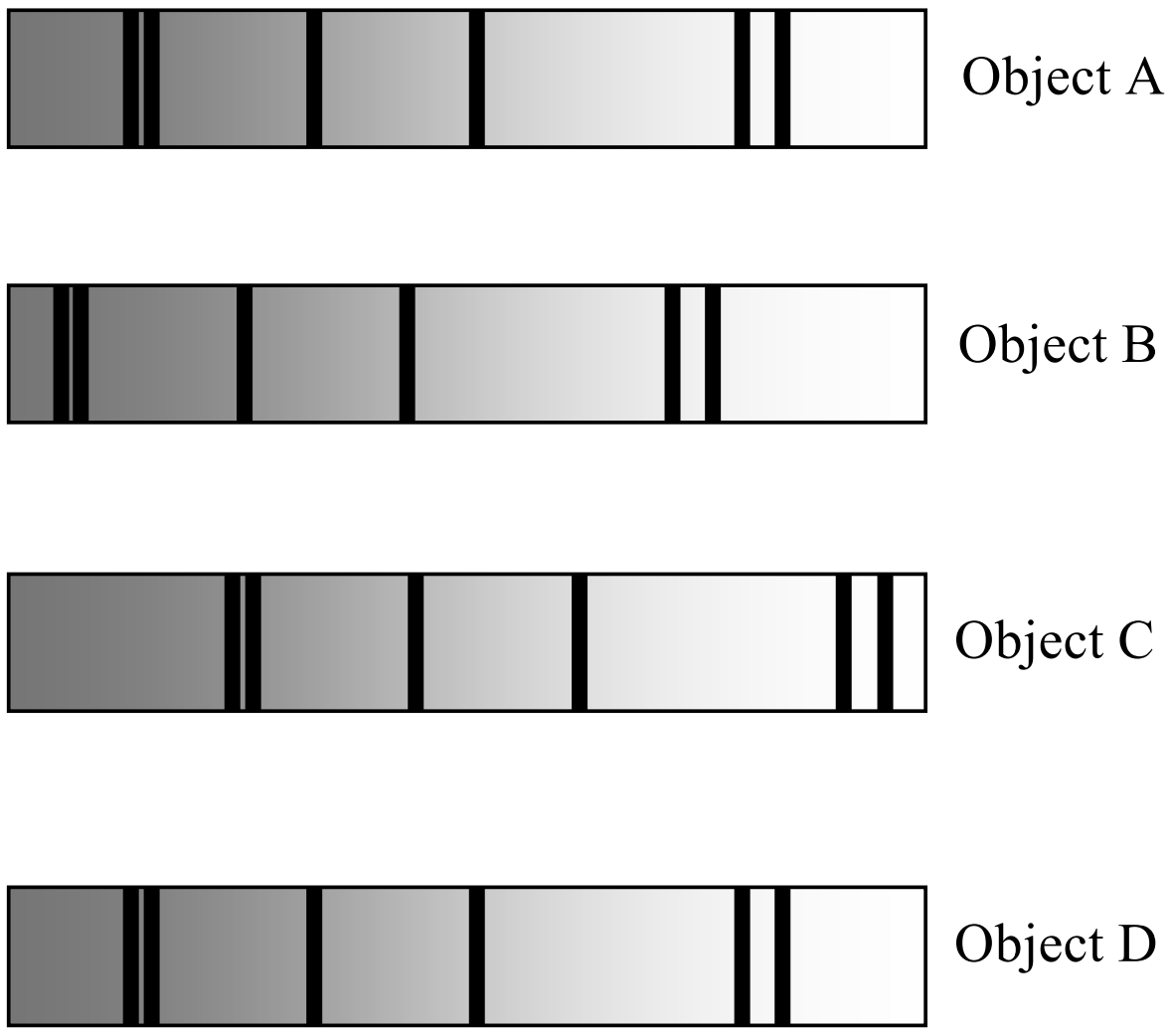

3. The figure below shows the orbit of a star-planet system. Given the location marked with the dot on the star's radial velocity curve, at what location (1-4) would you expect to find the planet at this time?
   a. 1
   b. 2
   c. 3
   d. 4

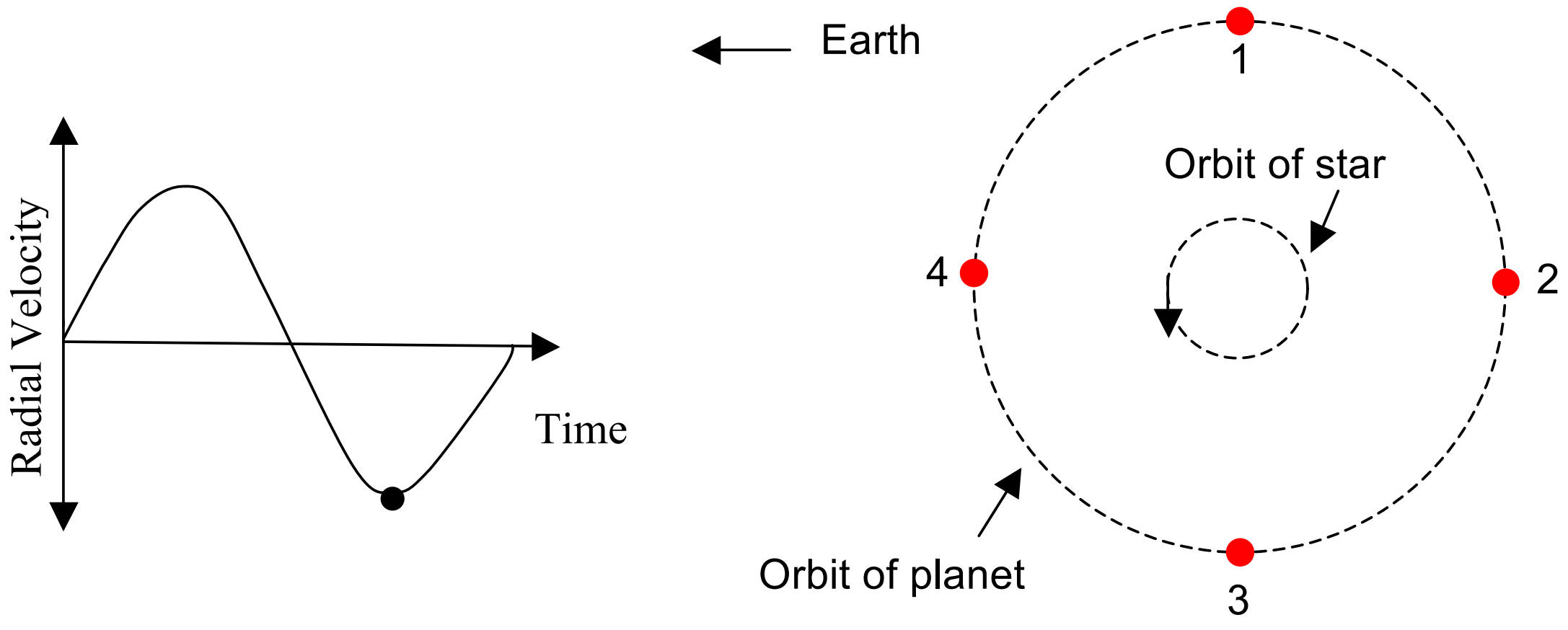

Figure 1. Three sample questions from the Astro 101 exams used in this study. All the questions require a grasp of one or more topics central to the course, and the second and third questions require multiple lines of reasoning to answer correctly. In the case of question 2, the student must note that all but one of the objects is blueshifted, the fourth being at rest, meaning that object C, having the most redshifted spectrum must be that object, and therefore object B, the most blueshifted, must be moving the fastest. In question 3, the student has to use the following line of reasoning to come to the correct answer: 1) the radial velocity curve shown is for the star, and the dot marks the point when the star has the most negative radial velocity; 2) since negative

velocity represents blueshifted motion, or motion towards the observer, this point corresponds to motion directly *towards* the Earth; 3) this moment in time corresponds to the point at the top of the star's orbit shown in the diagram; and 4) since the planet must always be on the opposite side of the orbit from the star (from the definition of center of mass), the planet must be at point 3 at that time.

## IV. ASSESSMENT OF STUDENT LEARNING: DID USING A TEXTBOOK MATTER?

Two methods of assessing student learning were used to assess the impact of making the textbook optional: the common exams used throughout the study, and research-validated concept inventories. First, results comparing student learning as measured by these two methods are presented, showing that there is no statistical difference in student learning whether the textbook is required or optional. Then, results of surveys of students' attitudes about the textbook are presented.

Table 2. Summary of exam scores

| | Spring 2007 | Fall 2007 | Spring 2008 | Fall 2008 | Spring 2009 | Fall 2009 | Spring 2010 | Fall 2010 | Spring 2011 | Weighted Average |
|---|---|---|---|---|---|---|---|---|---|---|
| | Textbook required | | | | | Textbook optional | | | | |
| Number of students | 60 | 56 | 59 | 57 | 108 | 60 | 107 | 59 | 101 | 667 |
| Midterm exam 1 | 74.6 | 69.5 | 73.3 | 68.9 | 76.9 | 73.3 | 71.6 | 77.1 | 76.2 | 73.8 |
| Midterm exam 2 | 76.1 | 72.7 | 75.4 | 75.8 | 74.6 | 73.6 | 73.0 | 77.3 | 72.0 | 74.2 |
| Final exam | 67.7 | 68.4 | 70.1 | 67.7 | 70.9 | 67.8 | 69.9 | 73.0 | 70.2 | 69.7 |

### 1. Exams

Table 2 lists the average exam scores for each of the nine times the course was taught, along with the overall average (weighted by the number of student in each class) for each exam for the entire period. A cursory examination of these exam scores shows no obvious change in scores before Spring 2009 and after Fall 2009, suggesting that making the textbook optional did not affect student learning. To test this proposition, detailed quantitative analysis of these data was performed. For all three exams, the weighted average and standard deviation were calculated for the two groups separately: those for whom the textbook was required (Spring 2007-Spring 2009) and those for whom it was optional (Fall 2009-Spring 2011). Figure 2 shows a plot of these data, giving support to the suggestion that there is no statistical difference between the exam scores of the two groups.

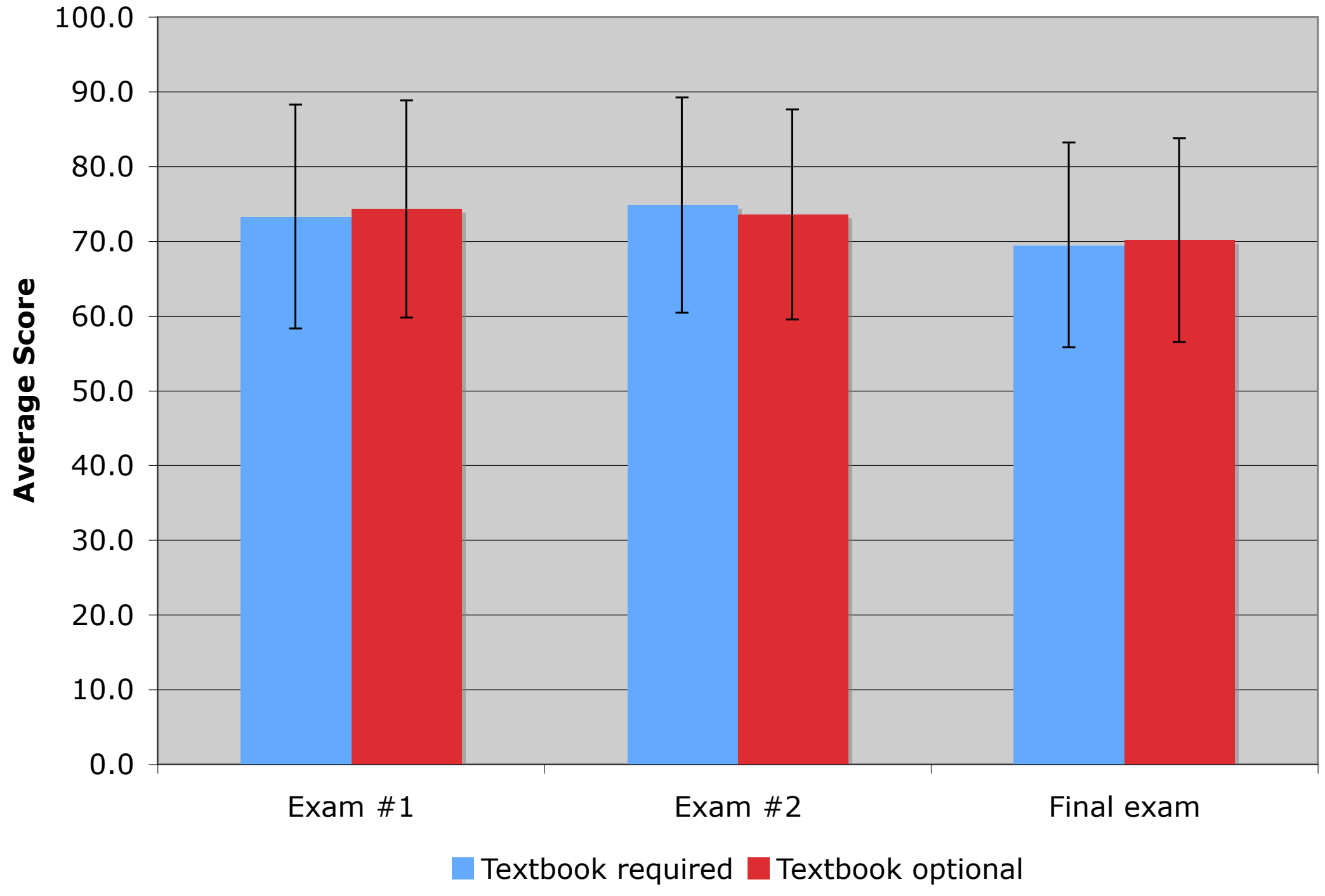

Figure 2. Plot of average exam scores for two groups: those for whom the textbook was a required purchase (blue bars; N=322-328) and those for whom the textbook was optional (red bars; N=319-325).[5] The error bars indicate the standard deviation of the samples.

To make a more quantitative estimate of whether making the textbook optional had any affect on students exam scores (and hence, presumably, on their learning), a 2-sample t-test assuming equal variance,[6] with $\alpha = 0.01$, was conducted comparing the scores on each exam for the two groups. The results of those t-tests strongly confirm the null hypothesis in all three cases: that there is no statistical difference between the two groups. The two-tail *p*-values of 0.367, 0.255, and 0.543 for the first midterm, second midterm, and final exam, respectively suggest that it is between 25-55% likely that any differences between the groups are due to statistical fluctuations rather than to any real differences between the groups. This finding does not mean that individual students did not benefit from using the book. However, these data suggest that, on average, students in this Astro 101 class were able to learn the material in the course as well without a textbook as with one.

[5] The variations in N are because a few students did not take each exam.

[6] The statistical equality of the variances for the two groups for each exam was confirmed using a 2-sample F-test for variances.

To further probe the effect of the textbook on student learning, the students who took the exams when the textbook was optional (Fall 2009-Spring 2011) were asked, using clickers, who had purchased a book during that period, and who had not. The weighted average and standard deviation for each exam were then calculated for each these two groups of students.

Figure 3 shows a plot of these data, again suggesting that there is no statistical difference between the exam scores of these two groups. A 2-sample t-test assuming equal variance,[7] with $\alpha = 0.01$, was conducted comparing the scores on each exam for these two groups. The results of those t-tests again strongly confirm the null hypothesis in all three cases: that there is no statistical difference between students who purchased a textbook and those who did not (two-tail *p*-values of 0.642, 0.900, and 0.618 for the first midterm, second midterm, and final exam, respectively).

[7] The statistical equality of the variances for the two groups for each exam was again confirmed using a 2-sample F-test for variances.

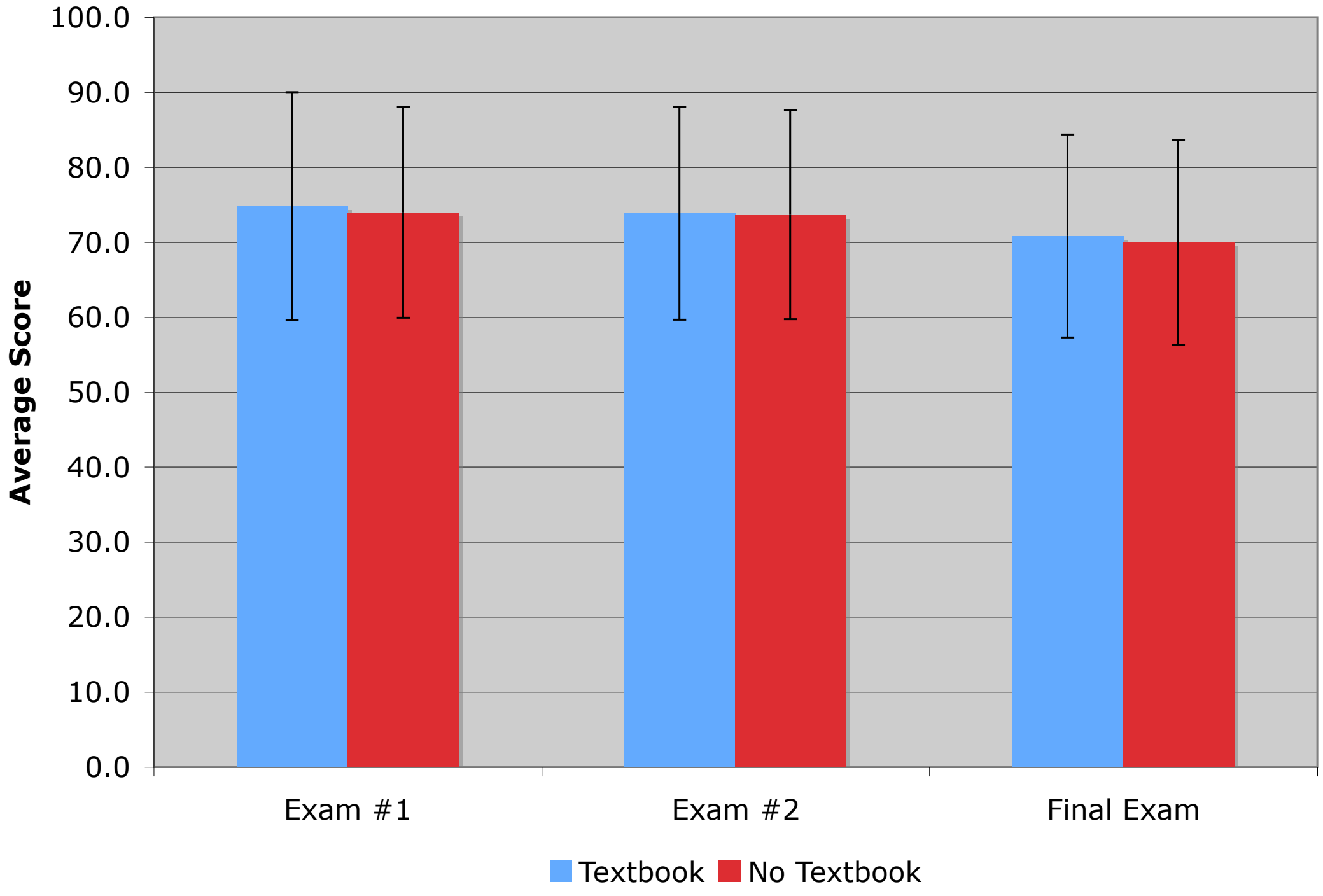

Figure 3. Plot of average exam scores for two groups: those who purchased the textbook (blue bars; N=119) and those who did not purchase the textbook (red bars; N=153), during the period when the purchase of the textbook was optional (Fall 2009-Spring 2011). The error bars indicate the standard deviations of the samples.

2. Concept Inventories

To further assess whether the use of the textbook had an impact on student learning, the Light and Spectroscopy Concept Inventory (LSCI; Bardar et al 2007) was used. The LSCI is a research-validated concept inventory designed to test students' conceptual knowledge of the concepts of light and spectra. A national study of student learning in Astro 101 classes using the LSCI administered to almost 4000 students in 69 classes at 31 universities showed a wide variation of normalized gain scores (0-50%), with the level of interactivity being the single most important variable in predicting the level of student learning in a class (Prather et al. 2009, Prather, Rudolph, & Brissenden 2009, Rudolph et al. 2010).

The LSCI was given, both pre- and post-instruction, to three of the Astro 101 classes when the textbook was required, and once when the textbook was optional. For each class, a class-averaged normalized gain was calculated for students who had taken both the pre-test and post-test. For the three classes with a required textbook, these gains were <*g*> = 0.47, 0.34, 0.34 ($N$ = 33, 28, 36), compared to the course with the optional textbook which had <*g*> = 0.38 ($N$ = 75). A 2-sample t-test assuming equal variance with $\alpha = 0.01$ comparing these two groups (required versus optional text) showed no statistical difference between these groups (two-tailed $p$ = 0.693). Thus, the LSCI results are consistent with, and reinforce the results from, the exam scores: student learning in this class did not depend on whether a textbook was required or not. A comparison of the demographic makeup of the classes (using the same demographics questions as in Rudolph et al. 2010) showed no statistical difference in the demographics of the students in the class before and after the textbook was made optional.

The students in the Spring 2011 class were also given the Star Properties Concept Inventory (SPCI; Bailey et al. 2011), to gauge their learning of another central set of concepts in an Astro 101 class. The class-averaged normalized gain was <*g*> = 0.41 ($N$ = 59), which compares favorably with the results of a national survey of SPCI scores, which had an average of <*g*> = 0.30 ($N$ = 334). Thus, although we cannot compare the results for the SPCI between the classes that did or did not have a required textbook, the excellent performance on the SPCI of the class with the optional textbook compared to a large, national sample of Astro 101 students (most or all of whom probably were using a textbook) is consistent with the above findings that a textbook is not required to learn concepts central to many Astro 101 courses, such as star properties.

3. Student attitudes

The students were surveyed, using clickers, numerous times, to determine their attitudes about the textbook. The first time was at the end of the Spring 2009 term, when the textbook was still required. When the students were asked if they would have preferred the book not be a required purchase, just over half (56%) of those surveyed said “yes”. The next four times the class was taught with an optional textbook (Fall 2009, Spring 2010, Fall 2010, and Spring 2011), the students were asked how many had bought the book. Overall, somewhat less than half (44%) of the students bought the textbook. The students were then asked how many were happy with their decision (whether to buy the book or not). Combining these results showed that two-thirds of the students (67%) either had not bought the book and were glad they did not, or had bought the book and were sorry they did: in other words, a large proportion of the class did not find the book useful or did not miss it. Of course, student opinion should not be the decisive factor in determining if a given instructional modality is working, but when reinforced by the quantitative data showing that the lack of a required textbook did not adversely affect student learning, the students’ opinions support the finding that the textbook did not add measurable value to their learning in a class structured and designed as this class was.

## V. DISCUSSION

The title of this article asks the question: Do you always need a textbook to teach Astro 101? The data suggest that the answer is: not necessarily. In the class studied here, students were able to perform as well on exams and learning assessments measuring student learning gains for some of the most fundamental topics in astronomy (e.g., light and spectra and stellar properties), whether a textbook was required or not. This result leads to the question of what material students can rely on to learn the material in a class without a required textbook. In the class studied here, the students had access to the Lecture-Tutorials and Ranking Tasks they had completed from which to learn, along with their class notes and the class Powerpoints [PPTs], which were available on-line after class.[8] For this class and the associated learning goals, these materials were sufficient to allow the students to learn the concepts of the course, as measured by the exams and independent assessments used. Of course, another instructor might choose different learning goals from the ones used in the class studied here, and might use different methods to ensure that students master those learning goals, including the use of a textbook. The results presented here do not rule out that such uses of a textbook can be useful to student learning. However, the proposition that a textbook is necessary to any course with learning goals that do not pertain directly to the use of the textbook (e.g., memorizing facts only presented in the textbook, and not in class), cannot be taken for granted, and should be subjected to the same type of research scrutiny as is presented here.

It is also essential that the assessment in the class closely match the instructional material. In the case of this course, the bulk of the exam materials used were developed using best research practices, independently of the course instructor. These questions probed the same concepts presented in the textbook as well as in the interactive learning materials used in the class (Think-Pair-Share questions, Lecture-Tutorials, Ranking Tasks). Some instructors worry that such matching is “teaching to the test.” “Teaching to the test” implies that students are taught only the information that will appear on the test, which is not the case here. In best practice, instructors make the learning objectives of the course clear, and then construct their assessments to test those learning objectives (Brissenden, Slater, & Mathieu 2002). Our assessment was crafted to match the learning goals of the course and the ideas, concepts, and skills taught in the course, and the questions on the assessments were designed to test students' higher-level learning of the material. Thus, we appropriately matched our expectations of the students to the tool we used to assess if they met those expectations, as good assessment should.

[8]The PPTs contained the Think-Pair-Share questions for each class (but not the answers), allowing students to use those questions as review.

One argument that a textbook could still be useful for student learning might be that the textbook used in this class was not a "refutational text" (Guzzetti et al. 1997), and therefore should not be expected to help students learn beyond the known limitations of traditional "non-refutational" texts. This point, while valid, is weakened by the fact that, to my knowledge, none of the large number of Astro 101 textbooks currently available is a "refutational" text, so as a practical matter, such a choice is not available.

New instructors may find a textbook helpful in organizing their course, and may appreciate knowing that students will be getting an alternative presentation of the material.[9] Also, studies of the efficacy of interactive learning strategies, such as those used in this study, have suggested that quality of implementation is critical to success with those strategies (Prather et al. 2009, Prather, Rudolph, & Brissenden 2009). Thus, one might argue that forgoing a textbook is better left until one is proficient with the alternative instructional materials and strategies.

---

[9] However, it is worth noting that treating the textbook as a "roadmap" may lead novice instructors to be over-inclusive in their topic list; modern textbooks are written to be comprehensive, so that instructors can design any type of course they like using it (e.g., stars, solar system, cosmology, survey, etc.).

Another concern might be whether we are doing our students a disservice by not requiring them to read a textbook: after all, the skill of reading difficult technical material and making sense of it is clearly an important college-level skill. In the recent book *Academically Adrift* (Arum & Roksa 2011), the authors document a decline in the amount of reading required of college students in recent decades, and show that students who take courses requiring them to write more than twenty pages per term and read more than forty pages per week show greater improvement in their learning as measured by the Collegiate Learning Assessment. However, it is worth noting that the course in the study presented here did require a great deal of technical reading comprehension, as well as the use of critical reasoning and writing, via the Lecture-Tutorials. Further, Arum & Roksa are looking at the whole college student, who is presumably taking multiple courses, which can emphasize different skills. Nonetheless, if an instructor considers reading a textbook to be a central course goal in its own right, then they should absolutely require it.

As we discover more about how students learn in the Astro 101 class, many instructors are questioning the importance of the textbook, particularly as they come to realize that much of the learning in a class can occur in the interactions between students, such as during the pair portion of a Think-Pair-Share question, or when the students are performing carefully crafted, research-validated activities designed to address and overcome their pre-conceptions, such as Lecture-Tutorials and Ranking Tasks. It is my hope that the results of this study will contribute to a robust and open discussion and debate about the use of textbooks in Astro 101 classes, allowing each instructor to make their own decision, informed by the data presented here.

ACKNOWLEDGEMENTS

I would like to acknowledge the help of David Consiglio of Bryn Mawr College for his invaluable statistical acumen. I would also like to thank Gina Brissenden and Ed Prather for their comments and help improving this paper, as well as for their trailblazing work developing and delivering the CAE Teaching Excellence Workshops that changed the way many of us teach Astro 101, and guided the design of the course presented here. Finally, I would like to thank Sidney Wolff, Tom Donnelly, and Maria Rudolph for their careful reading of the text and suggestions for improvement. This research was generously supported by the National Science Foundation under Grant No. AST-0847170, a PAARE Grant. Any opinions, findings, and conclusions or recommendations expressed in this material are those of the author and do not necessarily reflect the views of the National Science Foundation.

Appendix A

| Date | Reading List and Events | In-Class Activities |
|---|---|---|
| M, 9/28 | Chapter 1 – Our Place in the Universe | Size and Scale 1-2 (RT) |
| W, 9/30 | Chapter 3 – The Science of Astronomy | Kepler's 2nd Law (LT)<br>Kepler's 3rd Law (LT) |
| M, 10/5 | Chapter 4 – Making Sense of the Universe: Motion, Energy, and Gravity | Newton's Laws and Gravity (LT) |
| W, 10/7 | Chapter 5.1-5.3 – Light and Matter | Electromagnetic Spectrum (LT)<br>Light and Atoms (LT) |
| M, 10/12 | Chapter 5.4-5.5 – Light and Matter | Types of Spectra (LT)<br>Doppler Shift (LT) |
| W, 10/14 | **Exam #1** (Chaps. 1, 3, 4, 5)<br>Chapter S2 – Space and Time | |
| M, 10/19 | Chapter 6 – Telescopes | Telescopes and Earth's Atmosphere (LT) |
| W, 10/21 | Chapter 14.1-14.2 – Our Star | Sun Size (LT)<br>Parsec (LT) |
| M, 10/26 | Chapter 15.1 – Surveying the Stars | Luminosity, Temperature, & Size (LT)<br>Blackbody Radiation (LT) |
| W, 10/28 | Chapter 15.2-15.3 – Surveying the Stars<br>Chapter 16 – Star Birth | H-R Diagram (LT)<br>Star Formation and Lifetimes (LT) |
| M, 11/2 | Chapter 13 – Other Planetary Systems (not in your text – to be handed out) | Temperature and Formation of the Solar System (LT)<br>Motion of Extrasolar Planets (LT) |
| W, 11/4 | **Exam #2** (Chaps. 6, 13, 14, 15, 16, S2)<br>Chapter 17.1-17.2 – Star Stuff | |
| M, 11/9 | Chapter 17.3 – Star Stuff<br>Chapter 18.1-18.2 – The Bizarre Stellar Graveyard: White Dwarfs and Neutron Stars | Stellar Evolution (LT) |
| W, 11/11 | No Class (Veteran's Day) | |
| M, 11/16 | Chapter 18.3 – Black Holes<br>Chapter S3 – Spacetime and Gravity | |
| W, 11/18 | Chapter 19 – Our Galaxy<br>Chapter 20.1 – Types of Galaxies | Galaxy Classification (LT)<br>Milky Way Scales (LT) |
| M, 11/23 | Chapter 20.2-20.3 – Galaxies and the Foundation of Modern Cosmology | Expansion of the Universe (LT) |
| W, 11/25 | No Class (Day before Thanksgiving) | |
| M, 11/30 | Chapter 22 – Dark Matter, Dark Energy, and the Fate of the Universe | Dark Matter (LT) |
| W, 12/2 | Chapter 23 – The Beginning of Time<br>*Evaluations* | Looking at Distant Objects (LT)<br>Q&A |
| **M, 12/7 at 1:40 – 3:40pm in 8-4** | **Comprehensive Final Exam** | |